\documentclass[preprint,showpacs]{revtex4}
\usepackage{graphicx}
\begin{document}
\title{Loop Quantum Cosmology: Recent Progress}

\author{Martin Bojowald}
\affiliation{Max-Planck-Institut f\"ur Gravitationsphysik, Albert-Einstein-Institut, 14476 Potsdam, Germany}
\preprint{AEI--2004--017}
\begin{abstract}
Aspects of the full theory of loop quantum gravity can be studied
in a simpler context by reducing to symmetric models like cosmological
ones.  This leads to several applications where loop effects play a
significant role when one is sensitive to the quantum regime. As a
consequence, the structure of and the approach to classical
singularities are very different from general relativity: The quantum
theory is free of singularities, and there are new phenomenological
scenarios for the evolution of the very early universe including
inflation. We give an overview of the main effects, focussing on
recent results obtained by several different groups.
\end{abstract}

\maketitle

\section{Introduction}

One of the leading contenders for a quantum theory of gravity is loop
quantum gravity, which is a canonical quantization of general
relativity formulated in Ashtekar's variables. With its properties of
being background independent and non-perturbative it is well-equipped
for the understanding of extreme physical situations such as black
holes and the big bang. On the other hand, special techniques are
required which are essential for a well-defined framework but often
unfamiliar. These techniques and their implications can be illustrated
and tested in simple situations by introducing symmetries, which is
the origin of loop quantum cosmology. The symmetry reduction can be
done in such a way that the characteristic basic features of the full
theory survive in explicit models, which exhibit far-reaching
consequences even with contact to observations.

This overview starts with a brief introduction to loop quantum gravity
and its symmetric cosmological models. Early results of isotropic loop
quantum cosmology were the observation that the evolution equation
becomes discrete and non-singular, and that operators for inverse
powers of the volume are bounded with a modified behavior at small
scales (these aspects have been reviewed in more detail in
\cite{LoopCosRev}). The modified inverse volume is a non-perturbative
effect of the quantum theory, but it can nevertheless be studied in
effective classical systems which are easier to analyze. At the same
time, the quantization is subject to ambiguities leading to several
parameters in the resulting expressions which can be exploited in
phenomenological investigations.

The methods have been extended to anisotropic models and lead in
particular to a Bianchi IX dynamics which is not chaotic thanks to
small-scale modifications. The effective classical equations give an
intuitive illustration of the modified appeoach to the classical
singularity. Also the isotropic behavior is different at small scales
if matter is present, which can be illustrated in a bounce picture
replacing the big crunch. The same modified dynamics also has
consequences for inflationary scenarios with characteristic deviations
from the standard case at early stages. This can leave observable
imprints on the microwave background at large scales. Work in progress
includes a systematic derivation of perturbative corrections as well
as approaches to find the physical inner product and observables.

\section{Loop quantum gravity}

Loop quantum gravity \cite{Rov:Loops,ThomasRev} is based on Ashtekar's
variables \cite{AshVar,AshVarReell} which are a densitized triad
$E^a_i$ and the SU(2)-connection $A_a^i=\Gamma_a^i-\gamma K_a^i$ where
$\Gamma_a^i$ is the spin connection compatible with the triad and
$K_a^i$ the extrinsic curvature. The Barbero--Immirzi parameter
$\gamma$ can take any positive real value which does not influence the
classical theory. The main advantage of those variables for a
canonical quantization is that gravity now appears in the form of a
gauge theory with a compact gauge group (the group of triad rotations)
for which there are powerful background independent quantization
techniques. For Einstein's field equations additional constraints, the
diffeomorphism and the Hamiltonian constraint, have to be imposed.

To obtain a mathematically well-defined framework one chooses
appropriate basic objects which are constructed from the fields
$A_a^i$ and $E^a_i$ on a space manifold $\Sigma$. The basic ingredient
of a loop quantization \cite{LoopRep,ALMMT} is to use holonomies
$h_e(A)={\cal P}\exp\int_e A_a^i\dot{e}^a \tau_i dt\in \mbox{SU}(2)$
and fluxes $F_S(E)= \int_S E^a_in_a\tau^i d^2y$ where $\tau_i$ are
Pauli matrices, $\dot{e}^a$ is the tangent vector to the edge $e$ and
$n_a$ the co-normal to the surface $S$. If all curves $e$ and surfaces
$\Sigma$ are allowed, holonomies and fluxes contain the same
information as the original fields. For a quantization, however,
holonomies and fluxes are much better since they are smeared versions
of the fields obtained by natural integrations along curves and
surfaces. In contrast to a three-dimensional smearing, which is
usually employed in quantum field theory, the one- and two-dimensional
smearings can be done without introducing a background metric. This is
essential for a background independent quantization.

The quantum theory is then defined on a representation of the
holonomy-flux Poisson-* algebra, which also needs to be background
independent. It has recently been shown that such a representation is
unique \cite{FluxAlg,Meas,HolFluxRep,SuperSel}. States can most easily
be given in the connection representation as gauge invariant functions
of the connection via holonomies. A basis is given by spin network
states \cite{RS:Spinnet} which are associated with graphs in $\Sigma$
whose edges are labeled by irreducible SU(2)-representations. This
indicates that such a state depends on the connection only via the
holonomies along edges of the graph, each of which is taken in the
corresponding representation. The matrices are then contracted to a
gauge invariant function, which requires additional labels in vertices
if their valence is larger than three. This state space carries a
diffeomorphism invariant inner product, the Ashtekar--Lewandowski
inner product \cite{DiffGeom}. Basic operators are matrix elements of
holonomies which act by multiplication, and fluxes as derivative
operators which can be seen to count the intersection number of the
surface associated with the flux and the graph associated with the
state being acted on.

There are three characteristic features of this representation, the
hallmarks of any loop quantization:

\begin{enumerate}
 \item The Hilbert space before imposing the constraints is
 non-separable, owing to the fact that all spin network states with
 different graphs, which form a continuous set of labels, are
 orthogonal to each other.  \item Holonomies are well-defined
 operators by definition, but it is not possible to represent the
 connection $A_a^i$ itself as an operator.  \item Flux operators have
 discrete spectra as the intersection numbers above. This entails a
 discrete spatial geometry
\cite{AreaVol,Area,Vol2} since the
triad contains all information about the geometry of space.
\end{enumerate}

So far, only the basic operators have been described which now have to
be used to construct more complicated ones. In particular one has to
represent and solve the diffeomorphism and Hamiltonian constraint to
implement Einstein's equations. This can be done for the
diffeomorphism constraint \cite{ALMMT}, and there are well-defined
operators for the Hamiltonian constraint \cite{QSDI}. The latter,
however, are extremely complicated and not much is known about their
solution space. At this point it is advantageous to introduce
symmetries in order to apply the theory in a simpler context.

\section{Loop quantum cosmology}

Classically, symmetries are introduced simply by restricting to
symmetric basic fields. In the case of isotropy we have an isotropic
triad $E^a_i=p\Lambda_i^I X_I^a$ and connection
$A_a^i=c\Lambda_I^i\omega_a^I$ where $\omega^I$ and $X_I$ are
left-invariant (under the action of the symmetry group) 1-forms and
vector fields and $\Lambda$ is an SO(3)-matrix indicating the internal
su(2)-direction of the components. All these ingredients do not play a
physical role; the former are given once we choose the symmetry group
and the latter can be chosen freely and are pure gauge degrees of
freedom. The only physical components are $p$ and $c$ which are
related to the more familiar scale factor by the relations $|p|=a^2$
and $c=\frac{1}{2}(k-\gamma\dot{a})$ where $k$ can be zero for a flat
model or one for a closed model.

On these variables we have to impose the Hamiltonian constraint
\begin{equation} \label{constraint}
12\gamma^{-2}[c(c-k)+(1+\gamma^2)k^2/4]\sqrt{|p|}= 8\pi GH_{\rm
matter}(p,\phi,p_{\phi})
\end{equation}
where $G$ is the gravitational constant and $H_{\rm matter}$ the
matter Hamiltonian which depends on the metric and also on matter
fields and their momenta simply denoted as $\phi$ and
$p_{\phi}$. Using the relations to the scale factor one can see
immediately that the constraint is nothing but the Friedmann equation
\begin{equation}
 3(\dot{a}^2+k^2)a = 8\pi Ga^3\rho_{\rm matter}(a,\phi,p_{\phi})\,.
\end{equation}

Loop quantum cosmology is based on symmetric states in loop quantum
gravity which by definition are supported only on invariant
connections \cite{SymmRed,PhD}. Thus, all holonomies are of the form
\begin{equation}
 h_I=\exp\left(c\Lambda_I^i\tau_i\int\omega^I\right)= \cos(\mu
c/2)+2\Lambda_I^i\tau_i\sin(\mu c/2)
\end{equation}
where $\mu\in R$ is a real parameter for the edge length. Gauge
invariant isotropic states can only depend on the connection via those
holonomies such that they are of the form
$\psi(c)=\sum_{\mu}\psi_{\mu}e^{i\mu c/2}$ with a sum over a countable
subset of the real line. With the notation $\langle c|\mu\rangle:=
e^{i\mu c/2}$ the inner product is
$\langle\mu|\mu'\rangle=\delta_{\mu\mu'}$. Basic operators are the
holonomies $e^{i\mu'c/2}|\mu\rangle=|\mu+\mu'\rangle$ and the flux
$\hat{p}|\mu\rangle=\frac{1}{6}\gamma\ell_{\rm P}^2\mu|\mu\rangle$.

This representation has the same properties as observed before for
loop quantum gravity in general \cite{Bohr}: There is a non-separable
Hilbert space, only holonomies but not $c$ itself are promoted to
well-defined operators, and the flux $\hat{p}$ has a discrete
spectrum. It is not possible to derive a $c$-operator because the
holonomy operators are not continuous in $\mu$ at $\mu=0$ such that
the derivative does not exist. The flux has a discrete spectrum in the
sense that it has normalizable eigenstates $|\mu\rangle$, even though
the range of eigenvalues is continuous, the real line.

Comparing to a Wheeler--DeWitt quantization, on the other hand, shows
a completely different framework: the representations are
inequivalent. Furthermore, a Wheeler--DeWitt quantization would have a
well-defined operator for $c$ (which in this context is equivalent to
the extrinsic curvature), and a continuous spectrum of $p$ or the
scale factor $a$. Both these new features of the loop quantization
have significant further consequences. (One can obtain a similar
representation also with ADM variables, as done explicitly in
\cite{BohrADM}.)

\subsection{Discrete evolution}

Since there is no $c$-operator, the constraint (\ref{constraint}) has
to be quantized in an indirect way using holonomies. The resulting
quantized Friedmann equation becomes a difference equation which shows
drastic differences to the Wheeler--DeWitt equation at small
volume. As a consequence, cosmological singularities are absent, which
has been shown for all homogeneous models
\cite{Sing,IsoCosmo,Closed,HomCosmo,Spin}. As an example we discuss
the isotropic equation
\begin{eqnarray} \label{diff}
&& (V_{\mu+5}-V_{\mu+3})e^{ik}\psi_{\mu+4}(\phi)- (2+\gamma^2k^2)
(V_{\mu+1}-V_{\mu-1})\psi_{\mu}(\phi)\\\nonumber
&& + (V_{\mu-3}-V_{\mu-5})e^{-ik}\psi_{\mu-4}(\phi)
  = -\frac{4}{3}\pi
\gamma^3G\ell_{\rm P}^2\hat{H}_{\rm matter}(\mu)\psi_{\mu}(\phi)
\end{eqnarray}
for the wave function $\psi_{\mu}(\phi)$, where
$V_{\mu}=(\gamma\ell_{\rm P}^2|\mu|/6)^{3/2}$ are the volume
eigenvalues and $\hat{H}_{\rm matter}(\mu)$ the matter Hamiltonian
acting on the wave function.

As a recurrence relation, this equation does not break down when we
reach the classical singularity $\mu=0$, and it allows us to evolve
from positive $\mu$ to negative $\mu$ reaching a branch preceding the
classical singularity in internal time $\mu$ \cite{Sing}. Essential
for this property is the fact that the matter Hamiltonian satisfies
$\hat{H}_{\rm matter}(0)=0$ when it corresponds to
to quantized matter and geometry, e.g.\ $\hat{H}_{\rm
matter}=\frac{1}{2}\widehat{a^{-3}}\otimes\hat{p}_{\phi}^2+
\hat{a}^3\otimes W(\phi)$ for a scalar. This property will be shown
in the next subsection.

\subsection{Finite inverse scale factor operator}

For the matter Hamiltonian we need a quantization of $a^{-3}$, for
which we could try to use an inverse power of the flux
$\hat{p}$. However, since the flux has a discrete spectrum containing
zero, its inverse does not exist as a densely defined operator. There
is an alternative procedure to quantize $a^{-3}$ which comes
directly from the full theory \cite{QSDV}. One first rewrites the
object in a way which is better suited for quantization, e.g.\
$a^{-3}=(3(8\pi\gamma Gl)^{-1}\{c,|p|^l\})^{3/(2-2l)}$ where we do not
need any inverse power if $0<l<1$ (a comparison with the full theory
shows that some values for $l$ appear more naturally, e.g.\
$l=\frac{3}{4}$ if one uses the formulas of \cite{QSDV}; the resulting
expressions for this value have been used in recent applications
\cite{Inflation,InflationWMAP,BounceClosed}). Now the
connection component appears for which we do not have an
operator. This problem can again easily be avoided by using
holonomies, $a^{-3}= \left((4\pi\gamma Gl)^{-1}\sum_I{\rm
tr}(\Lambda_I^i\tau_i h_I\{h_I^{-1},|p|^l\})\right)^{3/(2-2l)}$. This
expression can now be quantized immediately by using holonomy and flux
operators and turning the Poisson bracket into a commutator. The
result is a well-defined, finite operator which does not show the
classical divergence of $a^{-3}$ at $a=0$ \cite{InvScale}. However,
since we had to reformulate the non-basic classical expression there
is room for several quantization ambiguities. One of them is the
parameter $l$ which, as we will see, has an effect on the
quantization. Another ambiguity comes from taking the trace, which can
be done in any SU(2)-representation, not just the fundamental one as
understood above.

\begin{figure}
\begin{center}
 \includegraphics[width=8cm]{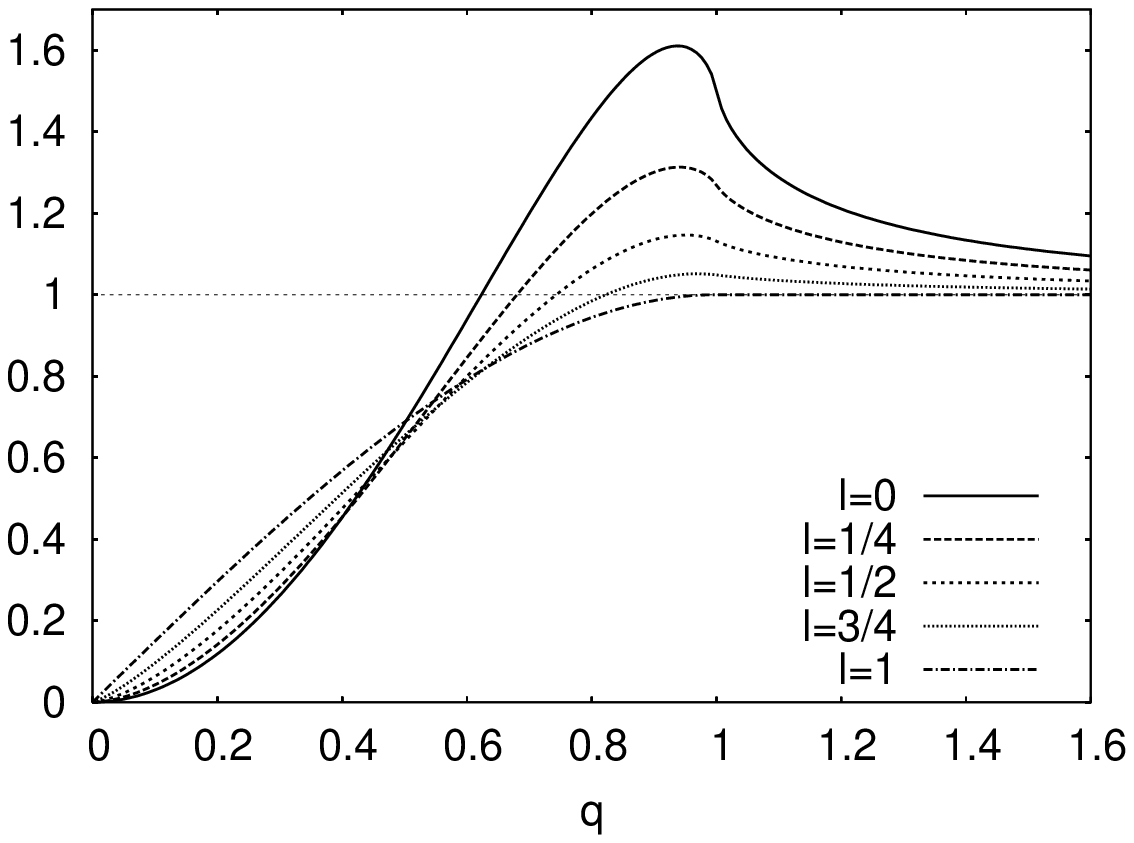} \includegraphics[width=8cm]{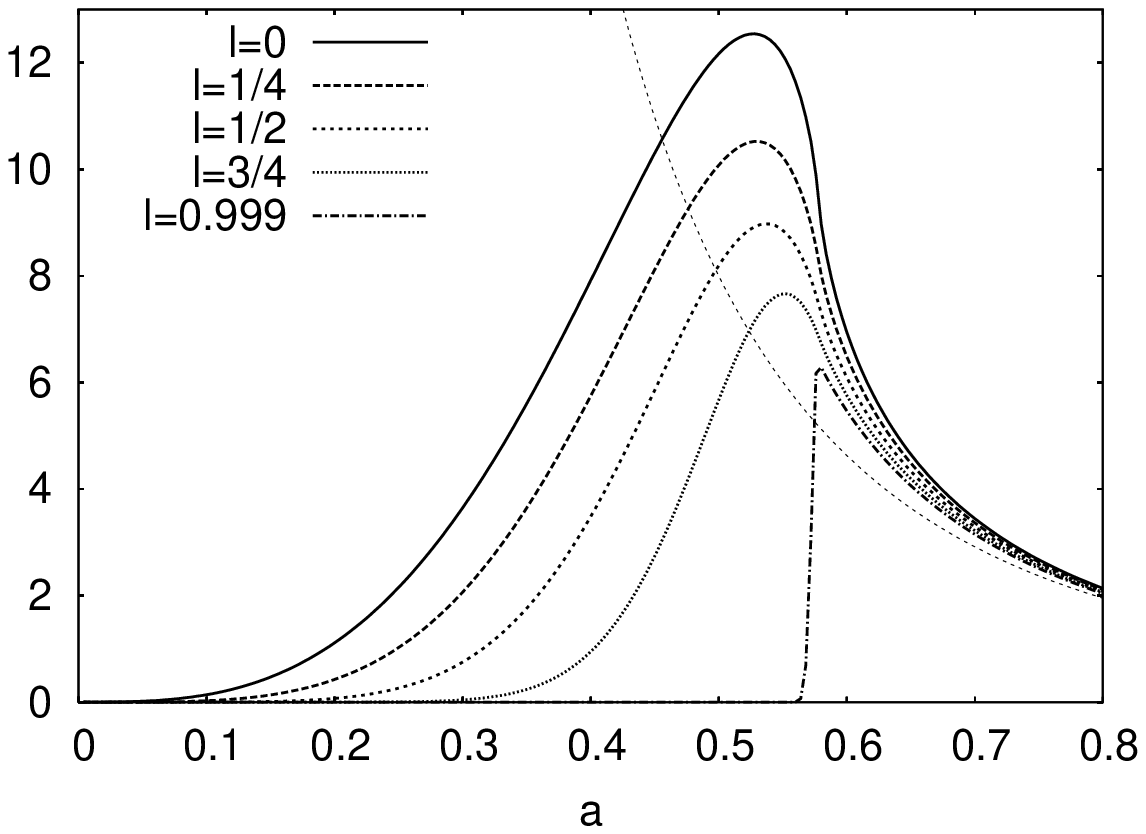}
\end{center}
\caption{The functions $p_l(q)$ and the corresponding effective
densities in units of $(\sqrt{\gamma j}\ell_{\rm P})^{-3}$, with $a$
in units of $\sqrt{\gamma j}\ell_{\rm P}$. Also shown are the
classical expectations one for $p_l(q)$ and $a^{-3}$ for the density
(dotted).}
\label{dens}
\end{figure}

If all these ambiguities are implemented, we obtain the eigenvalues
\begin{equation} \label{eigen}
 (\widehat{a^{-3}})_{\mu}^{(j,l)} = \left(\frac{9}{\gamma\ell_{\rm
P}^2lj(j+1)(2j+1)} \sum_{k=-j}^j k|p_{\mu+2k}|^l\right)^{\frac{3}{2-2l}}
\end{equation}
with $p_{\mu}=\frac{1}{6}\gamma\ell_{\rm P}^2\mu$.
For larger $j$, the sum can be approximated well by \cite{Ambig}
\begin{equation} \label{aeff}
 (a^{-3})^{(j,l)}_{\rm eff}(a):= (\widehat{a^{-3}})_{\mu(a^2)}^{(j,l)}=
 a^{-3} p_l(3a^2/\gamma j\ell_{\rm P}^2)^{3/(2-2l)}
\end{equation}
with $\mu(p)=6p/\gamma\ell_{\rm P}^2$ and
\begin{eqnarray} \label{pl}
 p_l(q) &=&
\frac{3}{2l}q^{1-l}\left((l+2)^{-1}
\left((q+1)^{l+2}-|q-1|^{l+2}\right)\right.\\
 &&\qquad- \left.(l+1)^{-1}q
\left((q+1)^{l+1}-{\rm sgn}(q-1)|q-1|^{l+1}\right)\right) \,. \nonumber
\end{eqnarray}
The functions $p_l(q)$ approach one for $q\gg1$ but are increasing
with power $p_l(q)\sim 3(l+1)^{-1}q^{2-l}$ for $q\ll1$. The transition
between the two regimes takes place around $q\approx 1$ (see
Fig.~\ref{dens}). For the effective density $(a^{-3})^{(j,l)}_{\rm eff}$
this implies that it does not diverge at $a=0$ but rather approaches
zero and is increasing as
\begin{equation}
 (a^{-3})_{\rm eff}^{(j,l)}\sim a^{3/(1-l)} \qquad\mbox{for
 }a\ll\sqrt{\gamma j}\ell_{\rm P}
\end{equation}
resulting in a power law behavior with some power larger than three at
small volume; see Fig.~\ref{dens}. (The power, however, is not
constant but decreases with increasing volume such that stability
issues like those in ordinary super-inflation are avoided.) Unlike the
classical reformulation of $a^{-3}$, the formula for the eigenvalues
can also be applied in the limit $l\to0$, which results in
\begin{equation}
 p_0(q):=\lim_{l\to0}p_l(q)=\frac{3}{2}q
\left(q+\frac{1}{2}(1-q^2)\log\frac{q+1}{|q-1|}\right)\,.
\end{equation}
In this limiting case, the effective density behaves just like the
volume at small scales.

The basic properties of the quantized $a^{-3}$ can be seen directly
from these expressions:
\begin{enumerate}
\item[$\mu=0$:] We have $(\widehat{a^{-3}})_{\mu=0}^{(j,l)}=0$ since in
this case we have a summation over an odd function in (\ref{eigen}). This
proves the claim made before that $\hat{H}_{\rm matter}(0)=0$ since
the metric operators in both the kinetic and the potential term
vanish.
\item[$0\leq\mu\leq 2j$:] $(a^{-3})^{(j,l)}_{\rm eff}(a)$ increases
with $a$, which implies a modified Friedmann dynamics. In particular,
the energy density increases with $a$ which implies inflation
\cite{Inflation}.
\item[$\mu\approx 2j$:] $(a^{-3})^{(j,l)}_{\rm eff}(a)$ has a peak where
the transition to the classical $a^{-3}$ takes place.
\end{enumerate}

This behavior is certainly very different from the one expected
classically, but it is very characteristic for the loop effects. In
fact, even though there are ambiguities like those indicated by the
parameters $(j,l)$, the features discussed here are robust. There are
more intuitive explanations for the effects which build on the
experience with quantum mechanics: The peak implies a finite upper
bound for $a^{-3}$ of the order $(\gamma j)^{-3/2}\ell_{\rm P}^{-3}$
which diverges in the classical limit $\ell_{\rm P}\to0$, just as the
finite ground state energy of the hydrogen atom,
$E_0=-\frac{1}{2}me^4/\hbar^2$ diverges in the classical limit.

Furthermore, the behavior around the peak provides an interpolation
between inflationary behavior at small scales (energy increasing with
$a$) and ordinary matter behavior at large volume which would
otherwise diverge when followed through to the classical
singularity. Thus, the energy density of matter in a universe behaves
differently when quantum effects are taken into account. A similar
situation is well-known for the energy density of radiation in a
cavity. In fact, it is analogous to the black body radiation whose
spectral energy density for large wave length, $\lambda\gg h/kT$, is
well described by the Rayleigh--Jeans law $\rho_T(\lambda)\sim 8\pi
kT/\lambda^2$ with an unphysical divergence at $\lambda=0$. Quantum
mechanics shows that the correct formula is given by Planck's
expression, $\rho_T(\lambda)=8\pi h\lambda^{-3}(e^{h/kT\lambda}-1)^{-1}
= h \lambda^{-3} f(\lambda/\lambda_{\rm max})$ with $\lambda_{\rm
max}=h/xkT$, $e^x(3-x)=3$, and $f(y)=8\pi/((3/(3-x))^{1/y}-1)$. The
peak position indicating the transition region is determined by the
inverse temperature $T^{-1}$.

Formally, the effective density in loop quantum cosmology behaves
similarly. The classical expression $a^{-3}$ is valid only at large
$a$, while the full expression is given by (\ref{aeff}),
$(a^{-3})^{(j,l)}_{\rm eff}(a)= a^{-3} p_l(a^2/a_{\rm
max}^2)^{3/(2-2l)}$, with a peak at $a_{\rm max}=\sqrt{\gamma
j/3}\,\ell_{\rm P}$. One could think that there is more than just a
formal analogy and that $j$, which determines the peak, is related to
an inverse temperature. In fact, if we choose the limiting case $l=0$,
we have $(a^{-3})_{\rm eff}^{(j,0)}\sim\Lambda_{\rm eff} a^3$ with
$\Lambda_{\rm eff}\propto a_{\rm max}^{-6}$. In this case, the
limiting behavior is asymptotically de Sitter at small $a$, which is
known to have a temperature $T=\sqrt{\Lambda}/2\sqrt{3}\pi$
\cite{GibbonsHawking}. With $\Lambda_{\rm eff}$, this becomes
$T\propto a_{\rm max}^{-3}\propto j^{-3/2}\ell_{\rm P}^{-3}$ such that
for the volume we have $V_{\rm max}/V\propto 1/VT$ and we can write
the effective density as $(a^{-3})_{\rm eff}^{(j,0)}(V)=V^{-1} g(VT)$
with a suitable function $g$, just as in the case of the black body
spectrum with $\lambda$ replacing $V$. At this point, however, there
is no indication that the ambiguity parameter $j$ can play such a
physical role analogous to the temperature.

\section{Recent applications}

Recently the original results of loop quantum cosmology have been
extended and studied in more detail, and more work is in
progress. The current research can mainly be grouped into the
following categories.

\subsection{Quantum structure of classical singularities}

For all homogeneous models that can be treated canonically (Bianchi
class A models) the absence of cosmological singularities can be shown
along the same general scheme as in the isotropic case
\cite{HomCosmo,Spin}. The quantum dynamics is given by a difference
equation which does not break down at the classical singularity but
instead extends to a branch beyond the classical singularity. The new
branch is always provided by triads of reversed orientation, just as
the sign of $\mu$ indicates the orientation in the isotropic case.

To get hints as for the behavior of general, inhomogeneous
singularities, one can make use of the BKL scenario \cite{BKL}. One
then views space approximately as being composed of almost homogeneous
patches each of which follows the most general homogeneous behavior:
the Bianchi IX model. Since this model is chaotic, patches with
initially similar geometries will depart rapidly from each other such
that all the patches will need to be subdivided to maintain the
approximation. Thus, the closer one comes to the classical singularity
the more fragmented space becomes without bound, which is not only
complicated but would also be inconsistent with a discrete structure
such as that of loop quantum gravity.

In this argument, however, only classical properties of the model have
been used. In particular the chaos comes from unbounded, diverging
curvature which leads to reflections in a complicated potential with
moving walls. From loop effects the curvature, just as the effective
density before, will be bounded and the walls will even disappear at
some finite volume \cite{NonChaos}. The complicated chaotic behavior
then stops at the discrete scale at the latest, making the loop
scenario consistent.

This argument is based on effective classical equations of motion
which have been modified by including the bounded curvature
expressions. Such a procedure is valid at reasonably large volume
where the classical trajectories describe the position of quantum wave
packets. Similarly, one can develop a classical picture of the
isotropic bounce. The classical equations are then modified by using
the effective matter Hamiltonian $\frac{1}{2}(a^{-3})_{\rm
eff}p_{\phi}^2+a^3 W(\phi)$ using the expression (\ref{aeff}). This
leads directly to the effective Friedmann equation
\begin{equation}
 3(\dot{a}^2+k^2)/a^2=8\pi G\left((a^3(a^{-3})_{\rm
 eff})^{-1}\dot{\phi}^2/2+W(\phi)\right)
\end{equation}
and, via Hamiltonian equations of motion, to the modified
Klein--Gordon equation \cite{Closed}
\begin{equation}
\ddot{\phi}=\dot{\phi}\,\dot{a}\frac{d\log(a^{-3})_{\rm
eff}}{da}-a^3(a^{-3})_{\rm eff}W'(\phi)\,.
\end{equation}

The main effect in the Klein--Gordon equation is that the friction
term can now change its sign in the transition region of
$(a^{-3})_{\rm eff}$ such that friction and antifriction for the
scalar will be interchanged. When we evolve toward a classical big
crunch, we have $\dot{a}<0$ such that classically with
$d\log(a^{-3})/da=-3/a<0$ we have antifriction. When we reach the
classical singularity, $\phi$ diverges as a consequence. With the
effective quantum behavior, however, $d\log(a^{-3})_{\rm eff}/da$ will
become positive once we reach small $a$. At this point, antifriction
turns into friction which leads to $\phi$ being frozen at a finite
value. This, in turn, implies that $\dot{\phi}\approx 0$ such that
only the potential term of the matter Hamiltonian is relevant and the
Friedmann equation gives a de Sitter bounce at non-zero $a$. This
scenario has been observed and studied for different potentials in
\cite{BounceClosed}.

After the bounce, we have $\dot{a}>0$ and classically friction for the
scalar, which is used in the slow-roll regime of inflation. With the
effective density, at small scales we have antifriction instead such
that $\phi$ will be driven away from the minima of its potential. This
has further consequences for cosmological phenomenology.

\subsection{Phenomenology}

We already mentioned that the increasing behavior of $(a^{-3})_{\rm
eff}$ at small volume implies, via the kinetic term of the matter
Hamiltonian, inflation \cite{Inflation}. There is no graceful exit
problem since inflation ends automatically once the scale factor
reaches the peak value of the effective density where it starts its
classical decrease. However, the usual cosmological perturbation
theory is unstable in this regime \cite{InflationWMAP} such that at
this point we cannot tell if this inflationary phase is viable for
structure formation. This phase will be present for any matter
content, even without an inflaton field. If we do couple an inflaton
with a potential suitable for chaotic inflation, there will be the
second effect of the inflaton being driven up its potential by the
antifriction term in its Klein--Gordon equation. Thus, starting from
its initial values, $\phi$ increases early before at later times it
reaches its maximal value which can be very large. This can provide
the initial conditions necessary for chaotic inflation
\cite{Inflation,InflationWMAP}. After reaching the maximum, $\phi$
turns around and enters the slow-roll phase of standard inflation.

For observational purposes it is important to notice that around the
turning point, where $\dot{\phi}\approx 0$, the slow-roll conditions
are violated \cite{InflationWMAP}. Since this happens at the earliest
stages of the slow-roll phase, it can lead to observable imprints such
as a suppression of power and a running spectral index at large scales
if the number of $e$-foldings is not too large.

\subsection{Perturbative corrections}

So far, most applications have been concerned with the modified
effective density since this is the most prominent effect. It is
non-perturbative and can extend even to large volume if the parameter
$j$ is chosen to be larger than one. Closer to the Planck scale there
are also perturbative corrections from the discrete structure of the
evolution equation \cite{SemiClass}. They can be computed
systematically from the expectation value of the Hamiltonian
constraint operator in a coherent state peaked at a phase space point
$(c_0,p_0)$. To leading order, the constraint equation then reduces to
the Friedmann equation, but there are correction terms of two
different kinds. In the flat isotropic model we have an asymptotic
expansion of the form \cite{Perturb}
\begin{equation}
 \langle \hat{H}\rangle =
-12\gamma^{-2}(c_0^2(1+O(c_0^2))+O(\ell_{\rm P}^2/p_0)+
O(d^{-2}))\sqrt{|p_0|}
\end{equation}
where the parameter $d$ characterizes the squeezing of the coherent
state. The first two correction terms are familiar and can be
interpreted as higher curvature corrections analogous to an effective
action. The last term is unexpected from this point of view; its
origin is the fact that there is no unique coherent state such as the
Minkowski vacuum for an effective action. The squeezing parameter $d$
can be constrained by further physical considerations for the
uncertainties of other quantities.

One may be tempted to relate this effective Hamiltonian to the
isotropic part of an effective action with higher curvature
corrections. In particular, local Lorentz invariance seems to be an
issue since there is no guarantee that the corrections with powers of
$c_0$, essentially the extrinsic curvature, will combine to the
correct combinations of the four-dimensional curvature
tensor. However, the relation between a perturbative effective
Hamiltonian and a perturbative effective action is more subtle since
higher curvature terms in the action imply higher derivatives. For
higher derivatives, in turn, the Legendre transform of the
perturbative action leads to a Hamiltonian which is not analytic in
the expansion parameter \cite{Simon}. From the canonical point of
view, however, only the analytic part of the Hamiltonian can be seen
in this perturbative analysis. Since only some terms of the curvature
invariants contain higher derivatives, they will be broken apart by
this procedure which can lead to apparent Lorentz violating terms in
the effective Hamiltonian. Note also that doing a perturbation
expansion does not commute with the Legendre transform for higher
derivative theories such that a derivation of an effective action from
the effective Hamiltonian is not straightforward.

\subsection{Conceptual issues}

One of the most important open issues conceptually is that of the
physical inner product and quantum observables. There are currently
several stategies which are being followed in cosmological models,
such as a Dirac analysis (see \cite{Golam} for a preliminary
investigation) and spin foam ideas.

This issue is also important in order to understand the number of
independent solutions. Even in the isotropic vacuum case, the
difference equation (\ref{diff}) has infinitely many solutions, all
but two of which are rapidly oscillating at small scales and do not
have a continuum approximation. The remaining two correspond to the
two independent solutions of the Wheeler--DeWitt equation. They are
further restricted by the quantum constraint resulting in only one
linear combination \cite{DynIn,Essay}. The meaning of the infinitely
many surplus solutions, however, and the possible role of
superselection can only be understood with the help of observables and
the physical inner product.

\section{Conclusions}

Loop quantum gravity takes the lesson of general relativity that the
metric is dynamical and should be quantized without artificial
background structures seriously. This requirement leads to a very
rigid structure and can only be realized with special mathematical
concepts. Chief among the basic mathematical properties are the fact
that only holonomies will become well-defined operators rather than
the connection or extrinsic curvature directly, and that the spatial
geometry via fluxes acquires discrete spectra. Both these properties
are also realized in loop quantum cosmology which allows reliable
tests of the full theory in this simpler context, and also an
investigation of further physical applications. As it turns out, the
consequences of these two basic aspects, and the most obvious
differences to a Wheeler-DeWitt quantization, are far-reaching.

The two properties entail immediately that the constraint (like other
composite operators) has to be quantized using holonomies, and that
the flux does not have an inverse operator such that its classical
inverse has to be quantized after using a reformulation. Physically,
this implies that the quantum evolution equation is a difference
rather than differential equation and that inverse powers of the scale
factor are modified at small scales, removing the classical
divergence. Both these effects are essential to show that the quantum
evolution is non-singular in all homogeneous models. On the other
hand, they also lead to corrections to the classical equations of
motion which are being used for phenomenological investigations. The
difference equation leads to perturbative corrections which are
relevant close to the Planck scale and so far have not been studied in
detail. The modification of inverse powers of the scale factor, on the
other hand, is non-perturbative and can extend into the semiclassical
regime. Consequences reach from new solutions for conceptual problems
to effects verging on observability.

It should be noted that all the modifications of the classical
equations of motion are not chosen with a particular cosmological
application in mind but rather derived from what we have learned about
crucial properties of a background independent, non-perturbative
quantization of general relativity. In all cases, the mathematical
properties were known to be essential before they had been seen to
lead to physical applications. In addition, there are now several
independent properties which need to play together in a special way in
order for the applications to hold true in the way described
here. Comparing with the Wheeler--DeWitt quantization, furthermore,
loop quantum cosmology has shown that inequivalent quantum
representations of the same classical theory can have essentially
identical behavior at large scales and at the same time lead to
drastically different conclusions when small-scale properties are
important.

\section*{Acknowledgements}

The author is grateful to B.~Iyer and the organizers of the
International Conference on Gravitation and Cosmology (ICGC 2004),
Cochin, India, for an invitation to a plenary talk and for providing a
stimulating atmosphere for further discussions.


\end{document}